\documentclass[preprint,12pt]{elsarticle}

\usepackage{multirow}
\usepackage{hyperref}
\usepackage{amstext}
\usepackage{graphicx}
\usepackage{epstopdf}
\usepackage{amssymb}
\usepackage{mathrsfs}
\usepackage{amsmath}
\usepackage{amsthm}
\usepackage{amsfonts}
\usepackage{verbatim}
\usepackage{color}
\usepackage{natbib}
\usepackage{stmaryrd}

\def\ang{\textup{\AA}}

\makeatletter
\newcommand{\myvec}[2][r]{%
  \gdef\@VORNE{1}
  \left(\hskip-\arraycolsep%
    \begin{array}{#1}\vekSp@lten{#2}\end{array}%
  \hskip-\arraycolsep\right)}

\def\vekSp@lten#1{\xvekSp@lten#1;vekL@stLine;}
\def\vekL@stLine{vekL@stLine}
\def\xvekSp@lten#1;{\def\temp{#1}%
  \ifx\temp\vekL@stLine
  \else
    \ifnum\@VORNE=1\gdef\@VORNE{0}
    \else\@arraycr\fi%
    #1%
    \expandafter\xvekSp@lten
  \fi}
\makeatother

\newtheorem{remark}{Remark}

\def\8{\infty}

\def\calA{\mathcal{A}}

\def\calN{\mathcal{N}}

\def\bfb{\mathbf{b}}
\def\bfF{\mathbf{F}}
\def\bfv{\mathbf{v}}
\def\bfx{\mathbf{x}}
\def\bfy{\mathbf{y}}

\def\bbR{\mathbb{R}}
\def\bbP{\mathbb{P}}
\def\bbQ{\mathbb{Q}}

\DeclareMathOperator*{\argmin}{arg\,min}

\journal{Journal of Computational Physics}

\begin{document}

\begin{frontmatter}

\title{Enriched Gradient Recovery for Interface Solutions of the Poisson-Boltzmann Equation}


\author{George Borleske, Y. C. Zhou\fnref{myfootnote}}
\address{Department of Mathematics, Colorado State University, Fort Collins, CO 80523}
\fntext[myfootnote]{Corresponding author. Email: yzhou@math.colostate.edu}



\begin{abstract}
Accurate calculation of electrostatic potential and gradient on the molecular surface is highly desirable for the continuum and hybrid modeling of 
large scale deformation of biomolecules in solvent. In this article a new numerical method is proposed to calculate these quantities on the dielectric 
interface from the numerical solutions of the Poisson-Boltzmann equation. Our method reconstructs a potential field locally in the least square sense
on the polynomial basis enriched with Green's functions, the latter characterize the Coulomb potential induced by charges near the position of reconstruction. 
This enrichment resembles the decomposition of electrostatic potential into singular Coulomb component and the regular reaction field in the Generalized Born methods. 
Numerical experiments demonstrate that the enrichment recovery produces drastically more accurate and stable potential gradients on molecular surfaces 
compared to classical recovery techniques.  
\end{abstract}


\end{frontmatter}


\section{Introduction}
The Poisson-Boltzmann theory has been widely accepted as a mean-field continuum approximation for electrostatic interactions 
in solvated biomolecular systems \cite{Honig1995_science}. The Poisson-Boltzmann theory treats the solute bimolecules as 
a singularly charged medium of low dielectric constant ($\epsilon_p = 1 \sim 2$) immersed in a high dielectric ($\epsilon_s = 75 \sim 80$) solvent with a 
continuum charge distribution that models the dispersed mobile ions. The large contrast of dielectric constant on
the highly complicated biomolecular surfaces poses a significant computational challenge and affords delicate mathematical and 
numerical treatments. The importance of Poisson-Boltzmann theory in biochemistry and biophysics has motivated 
extensive mathematical and computational investigations, see, for instance \cite{PBEQ,LuB2007e,RenP2012a,XieD2013a,ChenJ2018a,ZhongY2018a}, 
and references therein for the development of the subjects in the past decades and the latest overviews. 

This paper is concerned with the accurate calculation of the gradient of the electrostatic potential $\nabla \phi$ near the molecular surface 
from the numerical solutions of the Poisson-Boltzmann equation (PBE) for modeling solvated biomolecules. Traditionally the potential solutions 
of the Poisson-Boltzmann equation are used merely for computing the solvation free energy (and other derived energetic quantities such as pKa value, 
binding affinity, etc.), and therefore less attention had been paid on the calculation of the potential gradient. The success of Poisson-Boltzmann 
theory in these energetic evaluations promotes the exploration of PBE based electrostatic force calculations for molecular dynamics (MD) 
simulations \cite{MacKerellA1998a,LuQ2003a,Prabhu2004a,GengW2011a} and coarse-grained \cite{TangY2006a,MaL2009} 
or hybrid models \cite{ChenX2010a,ZhouY2010b,ChengL2013a}. There are three types of electrostatic forces defined by using the free energy 
functional derivatives and shape 
derivatives \cite{GilsonM1993a,ZhouY2008c,LiB2011a,MikuckiM2014a}, 
including (i) the body force exerted at each charged atoms of the solution molecules given by $qE = q \nabla \phi$ where $q$ is the local charge density; 
(ii) the dielectric boundary force exerted on the dielectric interface that is usually defined by the molecular surface; 
and (iii) the ionic pressure, exerted also on the molecular surface. 

Perpendicular to the dielectric interface, the dielectric boundary force 
\begin{equation}
f_n =-\frac{\epsilon_s}{2} |\nabla \phi^s|^2 + \frac{\epsilon_p}{2} |\nabla \phi^p|^2 - \epsilon_p |\nabla_n \phi^p|^2 + \epsilon_p (\nabla_n \phi^s) 
(\nabla_n \phi^p) \label{eqn:surface_force}
\end{equation}
poses significant challenges to the numerical computation because it is defined on the dielectric interface where one usually observes 
the peaks of the numerical error in the solutions of the 
Poisson-Boltzmann equation \cite{Stillinger1961a,Xiang1995a,Holst1995a,Holst2000a,BoschitschA2002a,Xie2007a,LinH2014a,Geng2017a}.
For boundary element methods, while $\nabla_n \phi^s$ can be obtained directly along with $\phi^s$ from well-conditioned boundary integral formulations, 
computing $\nabla \phi^p$ involves additional integrations of the numerical solution of the potential with a supersingular kernel over the entire 
surface \cite{LuB2006a,BajajC2011a,GengW2013a,ZhongY2018a}, usually causing a larger error. For finite difference interface methods, one usually extrapolates 
the potential solution across the interface by using the interface conditions to compute potential gradients \cite{GengW2011a}. For interface finite elements 
methods, gradient recovery techniques are recently developed so that potential gradient could be computed in a subdomain without using the solution values 
across the internal interface \cite{GuoH2018a,GuoH2017a,GuoH2018b}. 

We will develop an interface gradient recovery method to approximate the screened electrostatic potential from the numerical solution of the Poisson-Boltzmann 
equation. For elliptic problems with smooth coefficients, gradient recovery techniques have been well established on structured or unstructed 
grids \cite{ZienkiewiczO1992a,ZienkiewiczO1992b,BankR2003a,GuoH2015a,NagaA2005a,XuJ2003a,ZhangZ2005a}. By comparison, there is only a small handful of
numerical techniques concerning the gradient recovery from the primarily computed solutions of the elliptic problems with piecewise smooth 
coefficients or singular sources. For 1-D interface problems, special interpolation formulas are constructed to recover flux with high order accuracy 
from numerical solutions of linear and quadratic interface finite element methods \cite{ChouS2012a,ChouS2015a}. For 2-D elliptic interface problems on a 
specially constructed body-fitted mesh, gradient of the linear finite element solution is shown to be superclose to the gradient of the linear 
interpolation of the exact solution \cite{WeiH2014a}. By introducing the jump in the normal derivative as an augmented variable, Li {\it et. al.}
showed that a second-order convergence can be obtained for the solution and its gradient \cite{LiZ2017a}. Recently developed gradient recovery 
methods for interface problems are based on the second-order least square reconstruction of the solutions in an individual subdomain \cite{GuoH2018a}.
If the mesh is not conforming to the interface, a practice common in many immersed finite element methods \cite{LiZ1998a,LiuX2000a,HouS2013a,JiH2014a}, 
the solutions on the interface are first obtained on approperiately subdivided interface elements then supplied to the least square 
reconstruction \cite{GuoH2017a}. When Nitsche's method \cite{HanshoP2005a,AnnavarapuC2012a} is used for solving the elliptic interface problems, 
this subdivision of the interface element is not necessary, and one can use the solution on the extended fictitious subdomain to carry out the
least square reconstruction \cite{GuoH2018b}. Nevertheless, as we shall demonstrate below, these interface gradient recovery methods are not able 
to accurately compute the interface potential gradient on a spatial (domain or surface) discretization that are affordable for realistic 
biomolecular simulations.

Our method is based on the least square reconstruction of the numerical solution, with the classical polynomial basis supplemented with
$\{1/r_i\}$ where $r_i$ is the distance between a mesh node and the selected charges in the solute biomolecule. These additional functions have
been indicated in the Green's function for the Poisson equation and particularly in the generalized Born (GB) models for the biomolecular 
electrostatics. In GB models, the electrostatics potential at a charge $q_i$ inside the dielectric boundary is given by
\begin{equation}  
\phi(x_i) = \sum_{j \ne i} \frac{q_i q_j}{\epsilon_p| x_i - x_j|} + \bfF(x_i,x_j)  \label{eqn:model_GB}
\end{equation}
where $q_i$ is the singular charge at spatial position $\bfx_i$, and $\bfF$ is the reaction field due to the polarization charges induced at
the dielectric boundary \cite{ImW2003a,FeigM2004c}. A large variety of GB models have been developed to implicitly account for the effects of 
solvation through various parameterizations of effective Born radius, solvation accessible surfaces,  ionic screening, among others, see 
\cite{OnufrievA2019a} for a latest review and references therein. In the solvent domain $\Omega_s$, although an analytical approximation similar 
to (\ref{eqn:model_GB}) does not exist, the approximation 
\begin{equation}
\phi(x) = \sum_{i} \frac{q_i}{\epsilon_s| x - x_i|} e^{-\kappa |x - x_i|} \label{eqn:PB_bc}
\end{equation}
is widely used to compute the boundary conditions for the Poisson-Boltzmann at a boundary point $\bfx$ far away from the
solute molecules,  where $\kappa$ is the Debye-Huckel inverse screening length. The approximations (\ref{eqn:model_GB})-(\ref{eqn:PB_bc}) motivate 
us to choose functions of the form $1/r$ to enrich the classical polynomial basis function in the least square reconstruction. 
Numerical experiments demonstrate that our enriched recovery technique is highly accurate and stable for complicated biomolecules. 
By construction our recovery technique can be integrated with general interface numerical methods for the Poisson-Boltzmann equation, 
regardless of whether the discretization mesh is interface conforming or not.

The rest of the article is organized as follows. In Section \ref{sect:PBE}, the Poisson-Boltzmann equation and 
its numerical treatment are briefly reviewed, followed by the introduction of dielectric boundary force and its calculation.
We will present the our gradient recovery technique in Section \ref{sect:recover} and discuss its extension to enforcing the
interface conditions. Extensive numerical experiments will be conducted in Section \ref{sect:numeric} to verify the robustness
and accuracy of our methods on biomolecules of different complexity. The article concludes with a summary in 
Section \ref{sect:summary}.

\section{Poisson-Boltzmann Equation and Dielectric Boundary Force} \label{subsect:PBE}
This section presents the energetic theory of biomolecular electrostatics. The electrostatic free energy introduced by Sharp, Honig, Gilson {\it et. al.}
\cite{GilsonM1993a,GilsonM1985a,SharpK1990a,SharpK1990b} provides a unified theory to connect the electrostatic solvation energy and the Poisson-Boltzmann 
equation. The inconsistency between their definition of the electrostatic force and the sharp dielectric interface model was recently reconciled through
the introduction of a shape derivative of the solute biomolecules and the application of the Hadamard-Zol\'{e}sio structure 
theorem \cite{LiB2011a,MikuckiM2014a}. We will review the regularization techniques for the numerical solution of the PBE, which also sheds light to 
our choice of the enriching basis functions in the least square reconstruction for the gradient recovery.
\subsection{Biomolecular electrostatics theory and the Poisson-Boltzmann equation} \label{sect:PBE}
Consider $\Omega \in \bbR^3$ be a domain that encapsulates the solute biomolecules and the aqueous solution in which $M$ species of mobile ions disperse. 
The occupation domains of solute and solvent are respectively denoted by $\Omega_p$ and $\Omega_s$, with distinct dielectric constants $\epsilon_p$ 
and $\epsilon_s$. Denote by $\Gamma$ the dielectric interface separating these two subdomains. Assume dispersion of mobile ions follows the Boltzmann distribution, 
then the electrostatic free energy of this solvated system is given by
\begin{equation} \label{eqn:ele_eng}
G(\Gamma;\phi) = \int_{\Omega} \left[ -\frac{\epsilon}{2} | \nabla \phi|^2 + f \phi - \chi_s \beta^{-1} \sum_{j=1}^M c_j^0 (e^{-\beta q_j \phi} -1 ) \right] dx,
\end{equation}
where $c_j^0$ is the bulk concentration of $j^{th}$ species of ions, $f=\sum_{i=1}^N q_i$ is the collection of singular charges of biomolecule atoms, and
$\beta=1/(k_BT)$ is the inverse thermal energy at temperature $T$. The characteristic function $\chi_s=1$ in $\Omega_s$ and vanishes elsewhere. 
Equation (\ref{eqn:ele_eng}) highlights the dependence of the electrostatic free energy on the 
location of the dielectric interface $\Gamma$. Minimization of $G$ with respect to $\phi$ leads to the nonlinear Poisson-Boltzmann equation
\begin{equation} \label{eqn:PBE}
-\nabla \cdot( \epsilon \nabla \phi) + \chi_s \sum_{j=1}^M c_j^0 q_j  e^{-\beta q_j \phi} = \sum_{i=1}^N q_i \delta(x_i), 
\end{equation}
with interface conditions on $\Gamma$:
\begin{equation} \label{eqn:PBE_bc}
\left[ \phi(x) \right ] = 0, \quad \left[ \epsilon \nabla_n \phi \right] = 0,
\end{equation}
where $[\cdot]$ denotes the jump of the enclosed function and $\nabla_n$ is the normal derivative pointing to $\Omega_s$. 

To derive the dielectric boundary force we introduce a surface transformation $T_t$ corresponding to the slight change of 
domain $\Omega_p$ due to the motion of interface $\Gamma$ in its normal direction. The dielectric boundary force is derived by 
taking the variational derivative of the free energy $G$ with respect to  a velocity field $V$ normal to $\Gamma$. Associated with 
this velocity fields is a mapping $T_t(X)$ 
from the original surface coordinate $X \in \Gamma_0$ (material position) to the deformed coordinate $x \in \Gamma_t$ (physical position):
\begin{equation} \label{eqn:deformation}
T_t(X) = x(t,X) = x(0,X) + t \left . \frac{\partial x(t,X)}{\partial t} \right |_{t=0} + \mathcal{O}(t^2).
\end{equation}
Introducing the Jacobian $J_s$ of the surface transformation $T_t$:
\begin{equation} \label{eqn:J_deform}
J_s = \det \left(\nabla T_t(X) \right) \cdot | \nabla T_t^{-T} \cdot n(X)|,
\end{equation}
one can show that 
\begin{equation} \label{eqn:V}
\nabla \cdot V = \left. \frac{d J_s}{dt} \right |_{t=0}, 
\end{equation}
and the shape derivative of $G$
\begin{equation} \label{eqn:var_G}
\delta_{\Gamma} G(\Gamma; \phi) = \lim_{t \rightarrow 0} \frac{G(\Gamma_t;\phi) - G(\Gamma_0; \phi)}{t} = \int_{\Gamma_0} -f_n V \cdot n ds,
\end{equation}
where $f_n$ is the dielectric boundary force given in (\ref{eqn:surface_force}) \cite{MikuckiM2014a}.

Numerical techniques for the PBE are mostly focused on the treatment of the discontinuous dielectric function $\epsilon$ and the singular charge density 
$q_i \delta(x_i)$. There are three major regularization schemes with which a direct approximation of the singular delta functions can be avoided:
\begin{enumerate}[(I)]
\item  Subtract the potential 
\begin{equation} \label{eqn:phi_singular_1}
\phi^s(x) = \sum_{i=1}^N \frac{q_i}{\epsilon_s |x - x_i|}
\end{equation}
induced by the collection of singular charges from the potential in the entire domain $\Omega$, and thus one only needs to 
numerically solve the remaining regular potential $\phi^r = \phi - \phi^s$ \cite{ChenL2007a}. In solvent domain $\Omega_s$, this potential is much 
larger than the true potential $\phi = \phi^s + \phi^r$, so is the regular potential $\phi^r$ there. Consequently a small 
relative error in the numerical solution of $\phi^r$ will present a larger relative error in $\phi$, constituting an unstable
numerical algorithm \cite{HolstM2012a}. 
\item Subtract the singular potential (\ref{eqn:phi_singular_1}) only in the solute domain $\Omega_p$. This will create additional
jumps in the remaining regular components. Therefore a harmonic potential $\phi^h$ defined by
\begin{equation} \label{eqn:phi_harmonic}
\epsilon_p \Delta u^h = 0, ~ x \in \Omega_p; \qquad u^h(x) =-\phi^s, ~ x \in \Gamma,
\end{equation}
is added back to partially compensate the jump and to maintain $C^0$ continuity of the remaining regular potential at the interface $\Gamma$ \cite{ChernY2003a}. 
This strategy greatly improves the stability of the numerical method. On the other hand, one needs to compute $\nabla_n \phi^h$ on $\Gamma$
to supply the interface conditions for $\phi^r$:
\begin{equation} \label{eqn:PBE_bc_2}
\left[ \phi^r(x) \right ] = 0, \quad \left[ \epsilon \nabla_n \phi^r \right] = \epsilon_p \nabla_n \phi^h.
\end{equation}
Numerical evaluation of $\partial \phi^r/\partial n$ is nontrivial, and might introduce considerable error to the interface condition for $\phi^r$. 
A numerical Dirichlet-Neumann mapping based on boundary integral formulations will make this calculation accurate and efficient \cite{ZhouY2007b}.
\item One could choose not to add back the harmonic potential $\phi^h$ so there are only two components in the solute domain.
The absence of this harmonic component leads to the jumps in the potential and its normal derivative on the dielectric interface
\begin{equation} \label{eqn:PBE_bc_3}
\left[ \phi^r(x) \right ] =\phi^s(x), \quad \left[ \epsilon \nabla_n \phi^r \right] = \epsilon_p \nabla_n \phi^s.
\end{equation}
On the other hand, this regularization scheme helps reduce the complexity and improve the accuracy as one does not need to supply a numerical 
computed $\nabla_n \phi^h$ to the related interface condition for the regularized Poisson-Boltzmann equation \cite{Geng2017a}.
\end{enumerate}
Our enriched gradient recovery method is compatible with all these regularization schemes. We shall present the gradient recovery 
and force calculation along with the following general form of interface conditions:
\begin{equation} \label{eqn:PBE_bc_4}
\left[ \phi^r(x) \right ] = g(x), \quad \left[ \epsilon \nabla_n \phi^r \right] = h(x), 
\end{equation}
with known jumps $g(x),h(x)$.
\subsection{Dielectric boundary force and its calculation} \label{subsect:bc}
To get the dielectric boundary force $f_n$, one shall first compute $\phi_s, \phi_p$ and their respective gradients at selected positions (the centroids of
surface triangles, for example) on the interface $\Gamma$. There are there different procedures to complete this calculation:
\begin{enumerate}[(I)]
\item Independently compute $\phi_s,\phi_p$ and their gradients using proper gradient recovery techniques. 
\item Compute only one set of $(\phi_p, \nabla \phi_p)$ or $(\phi_s, \nabla \phi_s)$, and use the interface conditions (\ref{eqn:PBE_bc_4}) to get the other.
\item Compute both sets of $(\phi_p, \nabla \phi_p)$ and $(\phi_s, \nabla \phi_s)$ by enforcing the interface conditions (\ref{eqn:PBE_bc_4}) in the gradient
recovery.
\end{enumerate}
Procedure I is inferior to the others  because the resulting dielectric boundary force is not consistent with the interface conditions. 
Procedures II and III will be developed in Section \ref{sect:recover} and implemented in Section \ref{sect:numeric}

\section{Gradient Recovery Techniques Enriched with Green's Functions} \label{sect:recover}
Our gradient recovery techniques for PBE are based on the least square reconstruction of the numerical solution, which is also the 
basis of polynomial preserving recovery (PPR) techniques \cite{NagaA2005a,ZhangZ2005a} and their recent variations for elliptic 
interface problems \cite{GuoH2017a,GuoH2018a,GuoH2018b}. In this section we will first summarize the standard PPR method and 
then present its enrichment with Green's functions. This enriched gradient recovery technique will be integrated 
with different regularization schemes described in Section \ref{subsect:PBE} to enforce the respective interface conditions.

\subsection{Polynomial preserving recovery and enrichment with Green's functions}
We consider the molecular surface discretized by triangles and assume that the electrostatic potential $\phi$ has been solved from the 
Poisson-Boltzmann equation using a finite difference or finite element interface method. The mesh for the interface method may not be 
fitting the interface triangulation. The electrostatic force is usually computed at the centroids of triangles which are not 
necessarily the mesh nodes. Our method can be used to calculate the solution derivatives at vertices of surface triangles but one needs 
a delicate definition of the normal directions at the corner vertices to finally compute the surface force. For a centroid $o$ with 
coordinate $\bfx_o=(x_o,y_o.z_o)$ we define $\calN(o,n)$ be the collection of $n$ mesh nodes in solvent domain that are closest to $o$:
\begin{equation} \label{eqn:node_choice_1}
\calN_s(o,n) = \left \{v: v \in V_s, ~ |\bfx_v - \bfx_o| \le |\bfx_z-\bfx_o| \quad \forall z \notin \calN_s(o,n) \right \}, 
\end{equation}
where $V_s$ are the mesh nodes in $\Omega_s$ or on the molecular surface $\Gamma$. The numerical solution at this collection of nodes will
be fitted by a polynomial $p_s \in \mathbb{P}_{k}$ in the least square sense sampled at these nodes:
\begin{equation} \label{eqn:LS_1}
p_s(x) = \argmin_{p \in \bbP_{k}} \sum_{v \in \calN_s(o,n)} (\phi_s - p)^2(v).
\end{equation}
We then define the recovered gradient at surface position $o$ as
\begin{equation} \label{eqn:LS_2}
G_o(\phi_s) = \nabla p_s(o).
\end{equation}
Alternatively, one can choose the collection of nodes in the solute domain $\Omega_p$ to fit a polynomial $p_p(x)$ 
and to compute the recovered gradient 
\begin{equation} \label{eqn:LS_3}
G_o(\phi_p) = \nabla p_p(o).
\end{equation}
As in most analysis and applications of PPR, we choose $\bbP_{k}$ to be the space of quadratical polynomials, i.e., $k=2$.
\begin{remark}
The choice of sampling nodes (\ref{eqn:node_choice_1}) is for general interface methods. For finite element interface methods one
can take advantage of mesh structure to choose the mesh nodes in the first $l$ layers around the point $o$.
\end{remark}
\begin{figure}[!ht]
\begin{center}
\includegraphics[width=7cm]{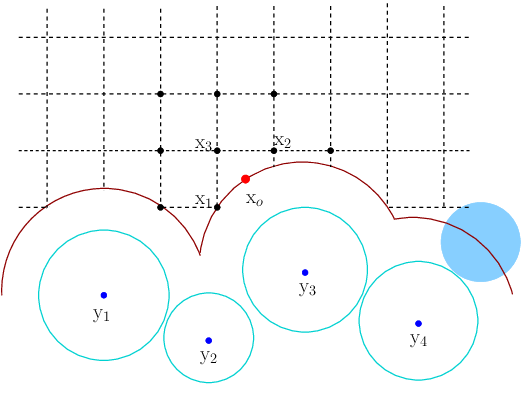} 
\caption{2-D illustration of polynomial preserving recovery (PPR) and enriched gradient recovery at a surface point $\mathbf{x}_o$ 
for the interface solution of the Poisson-Boltzmann equation on a Cartesian grid. PPR locally reconstructs solution using a polynomial 
in the least square sense at selected nodes $\mathbf{x}_i$ (black dots) in the solvent region $\Omega_s$ (meshed). Enriched recovery adds Green's 
functions induced by the charges $q_j$ at the centers $\mathbf{y}_j$ of selected atoms to the polynomial basis. The solvent accessible surface (SAS; red) 
is chosen to the dielectric interface. SAS is the trace of the center of probe sphere (blue) modeling water molecule as it rolls over the atoms (surface in cyan) 
of the solute molecule.} 
\label{fig:recovery}
\end{center}
\end{figure}

As we will demonstrated in Section \ref{sect:numeric}, the polynomial preserving recovery technique is not able to deliver an accurate surface
gradient on grid sizes affordable to practical biomolecular simulations, mostly because polynomial approximation of singular potential components
have large truncation errors. Since these singular components are induced by charges at known positions, we will introduce the Green's function
as basis functions in addition to the quadratical polynomials in the least square reconstruction, c.f. Fig. \ref{fig:recovery}. \textcolor{red}{We anticipate 
that the singular component of the solution will be approximated by these enriching basis functions, while the regular component of the solution will
be approximated in the original polynomial space.} For the surface point $o$ given above, we shall choose $m$ charged atoms of the solute molecule that 
are closest to $o$: 
\begin{equation} \label{eqn:atom_choose_1}
\calA_1(o,m) = \left \{a: a \in \calA_p, ~ |\bfy_a - \bfx_o| \le |\bfy_b-\bfx_o| \quad \forall b \notin \calA_1(o,m) \right \}, 
\end{equation}
where $\calA_p$ is the set of all charged atoms of the solute molecule. Alternatively, one may choose all charged atoms within a pre-determined 
distance $r_c$ to $o$:
\begin{equation} \label{eqn:atom_choose_2}
\calA_2(o,m) = \left \{a: a \in \calA_p, ~ |\bfy_a - \bfx_o| \le r_c \right \}. 
\end{equation}
\textcolor{red}{Since the singular components decay like $1/r$, contributions to the potential and force from charged atoms far away from $o$ are secondary,
and can be well approximated in the original polynomial space.}
By enriching the polynomial basis of degree $2$ with Green's functions centered at a total of $m$ selected charged atoms we define a new space of functions 
\begin{equation}
\bbQ_{2}  = \mathrm{span} \left \{ 1, x, y, z, x^2, y^2, z^2, xy, yz, xz, \frac{1}{|r_1|}, \cdots, \frac{1}{|r_{m}|} \right \}
\end{equation}
where $x,y,z$ are the Cartesian coordinates with respect to $o$ and $r_m$ is the distance to $m^{th}$ selected charge. We then 
fit the numerical solution at mesh nodes in $\calN_s(o,n)$ with a function $q_s \in \mathbb{Q}_{2}$ such that 
\begin{equation} \label{eqn:LS_4}
q_s(x) = \argmin_{q \in \bbQ_{2}} \sum_{v \in \calN_s(o,n)} (\phi_s - q)^2(v).
\end{equation}
This amounts to a least square solution $A \bfv = \bfb$ where each row of matrix $A \in \bbR^{n \times (10 + m)}$ 
is corresponding to a sampling mesh node in $\Omega_s$ with coordinate $(x,y,z)$ and is given by
\begin{align} 
A_{i,1:10} = & \left ( 1, \Delta x_i, \Delta y_i, \Delta z_i, \Delta x_i^2, \Delta y_i^2, \Delta z_i^2, \Delta x_i \Delta y_i, \Delta y_i \Delta z_i, 
\Delta x_i \Delta z_i \right ),  \label{eqn:Aij_1}   \\
A_{i,11:10+m} = & \left ( \frac{1}{|\bfx_i - \bfy_1|}, \cdots, \frac{1}{|\bfx_i - \bfy_{m}|} \right ), \label{eqn:Aij_2}
\end{align}
where $\Delta x_i = x_i - x_o, \Delta y_i = y_i - y_o, \Delta z_i = z_i - z_o$ and $\bfy_m$ is the position of the $m^{th}$ selected charge. 
Since the functions in (\ref{eqn:Aij_2}) are linearly independent
with each other and with the polynomial basis in (\ref{eqn:Aij_1}), the least square problem (\ref{eqn:LS_4}) is solvable.
\begin{remark} \label{remk:why_Greenfunc}
The potential in $\Omega_s$ does not have a decomposition with an analytical reaction 
field $\bfF(x,x_i)$ similar to (\ref{eqn:model_GB}). However,
the approximate solution (\ref{eqn:PB_bc}) suggests that it is possible to define a component of the form $1/r$. The space $\bbQ_2$ is constructed
to approximate this component using $1/r_i$ and the remaining component using the regular polynomial basis.
\end{remark}
\begin{remark} \label{remk:Q2_derivative}
The derivative of the Green's funtion can also be written in terms of the same Green's function. Thus the derivative of the basis functions 
in $\bbQ_2$ is also in $\bbQ_2$, allowing us to approximate both the potential and its derivatives in the same space.
\end{remark}
\begin{remark} \label{remk:Q2_bornion}
The approximation of the electrostatic potential in $\bbQ_2$ for the Born ion \cite{RouxB1990a} is exact.
\end{remark}

\subsection{Coupled gradient recovery with interface condition enforcement}
We need potentials and potential gradients on both sides of the dielectric interface to compute the dielectric boundary force. 
These quantities independently recovered on either side using the method described above do not satisfy the 
interface conditions, although these conditions have been enforced in the interface methods for the numerical solution 
of the PBE. Procedure II (c.f. Section (\ref{subsect:bc})) recovers the solution on one side and generate the surface potential and 
surface gradient on the other side using the interface conditions. This requires only one least square solution for the gradient recovery
at an interface point $o$, and thus is simpler to implement. It is worth noting that if one chooses to 
compute $(\phi_p, \nabla \phi_p)$ instead of $(\phi_s, \nabla \phi_s)$, then the enrichment using Green's function is not necessary, 
because the singular potential component has been removed from $\Omega_p$ in all three regularization schemes here. Coupled gradient recovery
on both sides of the interface is also possible by enforcing the interface conditions in least square reconstruction.
To this end we shall define in addition to $\calN_s(o,n)$ another collection of $n$ mesh nodes in the solute domain, $\calN_p(o,n)$, that are closest to $o$:
\begin{equation}
\calN_p(o,n) = \left \{v: v \in V_p, ~ |\bfx_v - \bfx_o| \le |\bfx_z-\bfx_o| \quad \forall z \notin \calN_p(o,n) \right \}.
\end{equation}
Two functions, one polynomial $p_p \in \bbP_2$ and the other $q_s \in \bbQ_2$, respectively sampled on $\calN_p(o,n)$ and $\calN_s(o,n)$, will be sought
from the following problem:
\begin{equation} \label{eqn:LS_5}
(p_p(x),q_s(x)) = \argmin_{p \in \bbP_2, q \in \bbQ_{2}} \left [ \sum_{v \in \calN_p(o,n)} (\phi_p - p)^2(v) + \sum_{u \in \calN_s(o,n)} (\phi_s - q)^2(u) \right ],
\end{equation}
subject to the interface conditions:
\begin{equation} \label{eqn:LS_interfacecondition}
q_s(o) - p_p(o) = g(o), \quad \epsilon_s \nabla_n q_s(o) - \epsilon_p \nabla_n p_p(o) = h(o).
\end{equation}
The solution of (\ref{eqn:LS_5}) leads to a least square problem $A \bfv = \bfb$ where the matrix $A \in \bbR^{(2n+2) \times (20+m)}$ whose first $n$ rows correspond
to the approximation of the solution $\phi_p$ respectively at $n$ sampling nodes $(x_i^p, y_i^p, z_i^p)$ in $\Omega_p$, i.e., 
\begin{align} 
A_{i,1:10} = & \left ( 1, \Delta x^p_i, \Delta y^p_i, \Delta z^p_i, (\Delta x^p_i)^2, (\Delta y^p_i)^2, (\Delta z^p_i)^2, \Delta x^p_i \Delta y^p_i, 
\Delta y^p_i \Delta z^p_i, \Delta x^p_i \Delta z^p_i \right ),  \label{eqn:Aij_3}   \\
A_{i,11:20+m} = & \left ( 0, \cdots, 0 \right ), \label{eqn:Aij_4} 
\end{align}
for $1 \le i\le n$. The next $n$ rows of $A$ are for the approximation at sampling nodes $(x_i^s, y_i^s, z_i^s)$ in $\Omega_s$, i.e.,
\begin{align} 
A_{n+i,1:10} = & \left ( 0, \cdots, 0 \right ), \label{eqn:Aij_5} \\
A_{n+i,11:20} = & \left ( 1, \Delta x^s_i, \Delta y^s_i, \Delta z^s_i, (\Delta x^s_i)^2, (\Delta y^s_i)^2, (\Delta z^s_i)^2, 
\Delta x^s_i \Delta y^s_i, \Delta y^s_i \Delta z^s_i, \Delta x^s_i \Delta z^s_i \right ),  \label{eqn:Aij_6}   \\
A_{n+i,21:20+m} = & \left ( \frac{1}{|\bfx_i - \bfy_1|}, \cdots, \frac{1}{|\bfx_i - \bfy_{m}|} \right ), \label{eqn:Aij_7}
\end{align}
for $1 \le i \le n$. The last two rows of $A$ come from the approximation of the interface conditions (\ref{eqn:LS_interfacecondition}): 
\begin{eqnarray} 
A_{2n+1,1}  & = &-1, \label{eqn:Aij_7a} \\
A_{2n+1,11} & = & 1, \label{eqn:Aij_7b}  \\
A_{2n-1,(2:10,12:20)} & = & (0, \cdots, 0), \label{eqn:Aij_8}   \\
A_{2n+1,21:20+m} & = & \left ( \frac{1}{|\bfx_o - \bfy_1|}, \cdots, \frac{1}{|\bfx_o - \bfy_{m}|} \right ). \label{eqn:Aij_9} \\
A_{2n+2,(1,5:10,11,15:20)} &  = &(0,\cdots,0), \label{eqn:Aij_10}   \\
A_{2n+2,2:4} & = & -\epsilon_p (n_x, n_y, n_z), \label{eqn:Aij_10a} \\
A_{2n+2,12:14}  & =  & \epsilon_s(n_x, n_y, n_z), \label{eqn:Aij_11}   \\
A_{2n+2,21:20+m} & = & \epsilon_s \left ( \frac{(\bfx_o - \bfy_1) \cdot n}{|\bfx_o - \bfy_1|^3}, \cdots, \frac{(\bfx_o - \bfy_m) \cdot n}{|\bfx_o - \bfy_{m}|^3} \right ). 
\label{eqn:Aij_12}
\end{eqnarray}
where $n=(n_x, n_y, n_z)$ is the unit outer normal on the interface. Solution of this least square problem reconstructs functions that 
approximate the potential solutions at sampling mesh nodes on two sides of the interface and the jump conditions at the interface point $o$.
 
\section{Numerical Experiments} \label{sect:numeric}
In this section we validate the accuracy and stability of our enriched gradient recovery method on a set of biomolecules. 
Since there do not exist exact solutions of the electrostatic surface potential and gradient for general molecules with 
complex surface geometries, we make up potential
fields for accuracy validation of Procedure II by setting the potential in the solvent domain $\Omega_s$ to be
\begin{equation} \label{eqn:exact_field}
\phi_s(\bfx) = \sum_{i=1}^N \frac{q_i}{\epsilon_s| \bfx - \bfy_i|}, 
\end{equation}
following the approximation boundary condition (\ref{eqn:PB_bc}). The exponential screen term is neglected because the ion strength $\kappa$ 
is alway small and makes very slight changes to the surface potential and gradient. Procedure II will be used in enriched recovery with analytical fields.
We will compare the classical polynomial preserving (PPR) and enriched gradient recoveries on molecules with the potential field (\ref{eqn:exact_field}). 
We will further solve the Poisson-Boltzmann equation with an interface finite difference method MIB \cite{ZhouY2006a,GengW2007a}, and use the solution at a 
fine mesh as the reference to check the accuracy of both PPR and our enriched gradient recoveries. Procedure III will be used since we have solutions 
available on both sides of the interface. We will use the following absolute and relative errors for quantifying the 
accuracy of recovered surface potential $R(\phi)$, gradient $R(\nabla_n \phi)$, and force $R(f_n)$ on the interface:
$$ e(\phi) = \| R(\phi) - \phi \|_{0, \Gamma}, e(\nabla_n \phi) = \| R(\nabla_n \phi) - \nabla_n \phi \|_{0, \Gamma},  
e(f_n) = \| R(f_n) - f_n \|_{0, \Gamma}, $$   
$$e_r(\phi) = \frac{e(\phi)}{\| \phi \|_{0, \Gamma}}, e_r(\nabla_n \phi) = \frac{e(\nabla_n \phi)}{\| \nabla_n \phi \|_{0, \Gamma}}, 
e_r(f_n) = \frac{e(f_n)}{\|f_n \|_{0, \Gamma}}.$$ 

\subsubsection{Born ion} \label{sect:born}
\begin{figure}[!ht]
\begin{center}
\includegraphics[height=4.5cm]{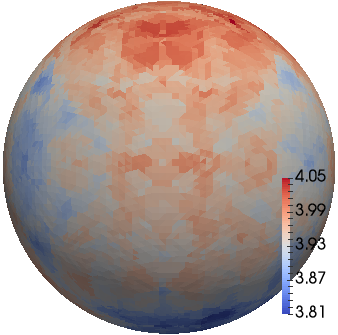} \hspace{1cm}
\includegraphics[height=4.5cm]{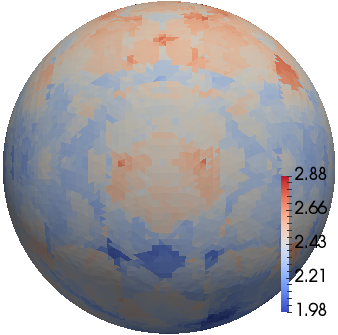} 
\caption{Surface potential (left; exact value $4.15$ uniform) and magnitude of the surface gradient (right; exact value $4.15$ uniform) recovered using 
PPR for Born ion. Mesh size $h=0.4$\AA.}
\label{fig:1atm_exact_nor}
\end{center}
\end{figure}

The potential field (\ref{eqn:exact_field}) is exact for Born ion so our enriched gradient recovery method will give result with machine error, 
for that $1/| \bfx - \bfx_i|$ is one of the basis function of $\bbQ_2$. For comparison, we present in Table (\ref{tab:1atm_nor})
the errors in surface potential and gradients generated by PPR. It is seen that PPR has proved optimal rate of 
convergence and can give fairly good results when the mesh is sufficiently refined. In particular, a mesh space smaller 
than $0.1$\ang~is needed to generate a surface potential gradient with a relative error less than $10\%$, an accuracy barely 
sufficient for the real biomolecular simulations. The surface potential and magnitude of the surface gradient recovered using 
PPR at $h=0.4$\AA~ are shown in Fig. \ref{fig:1atm_exact_nor}.  
\begin{table}[!ht]
\begin{center}
	\begin{tabular}{ r llll } \hline
		$h$ & $e(\phi)$  & $e_r(\phi)$ & $e(\nabla \phi)$ & $e_r(\nabla \phi)$ \\ \hline
		0.4  & 1.33 & 9.95e-2 & 7.97 & 1.18 \\ 
		0.2  & 2.10e-1 & 1.45e-2 & 3.02 & 2.58e-1 \\ 
		0.1  & 4.15e-2 & 2.83e-3 & 1.05 & 7.69e-2 \\ 
		0.05 & 6.69e-3 & 4.55e-4 & 3.16e-1 & 2.19e-2 \\ \hline
	\end{tabular}
	\caption{Errors and relative errors in surface potential and gradient recovered using PPR from the potential field (\ref{eqn:exact_field}) in $\Omega_s$
for Born ion.}
\label{tab:1atm_nor}
\end{center}
\end{table}

\subsubsection{Diatomic model} \label{sect:diatomic}
\begin{figure}[!ht]
\begin{center}
\includegraphics[height=3cm]{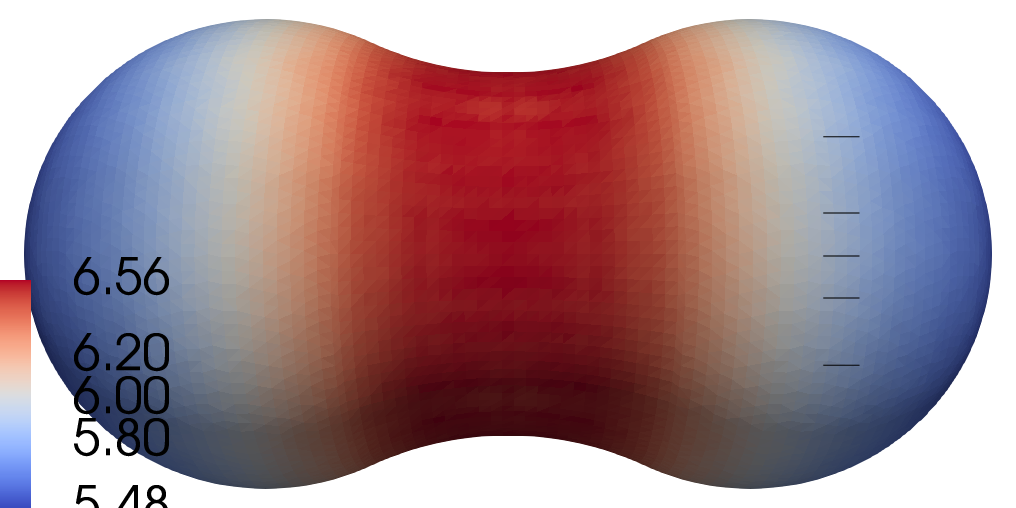} \hspace{1cm}
\includegraphics[height=3cm]{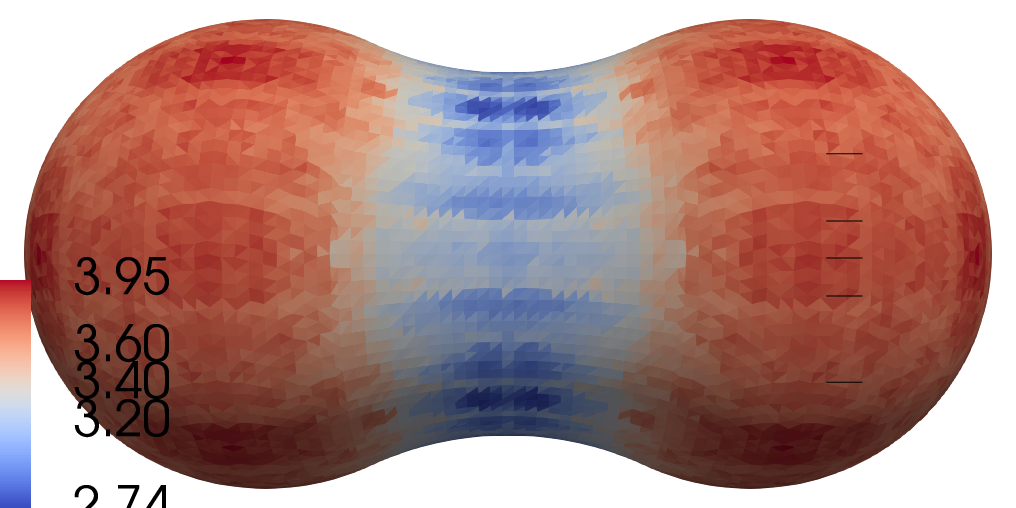}
\vspace{2mm}
\includegraphics[height=3cm]{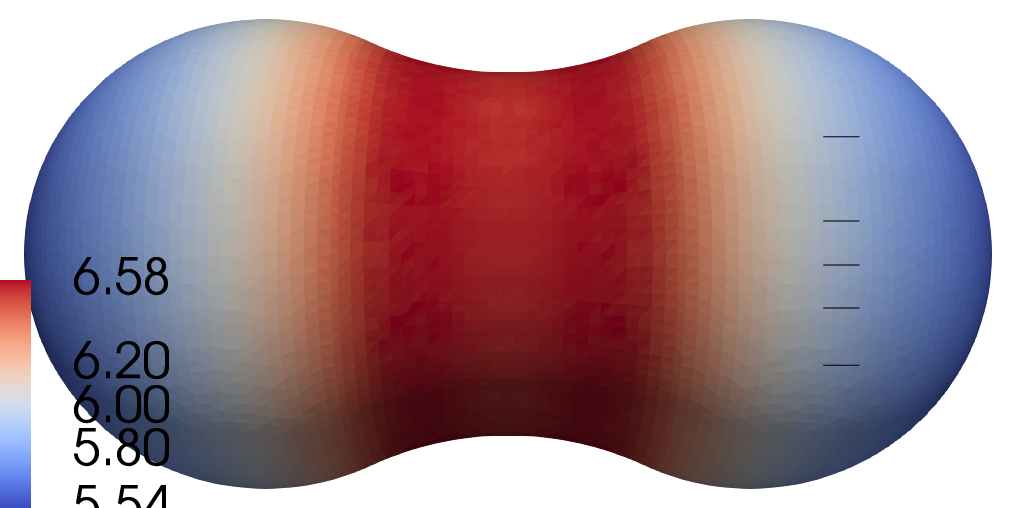} \hspace{1cm}
\includegraphics[height=3cm]{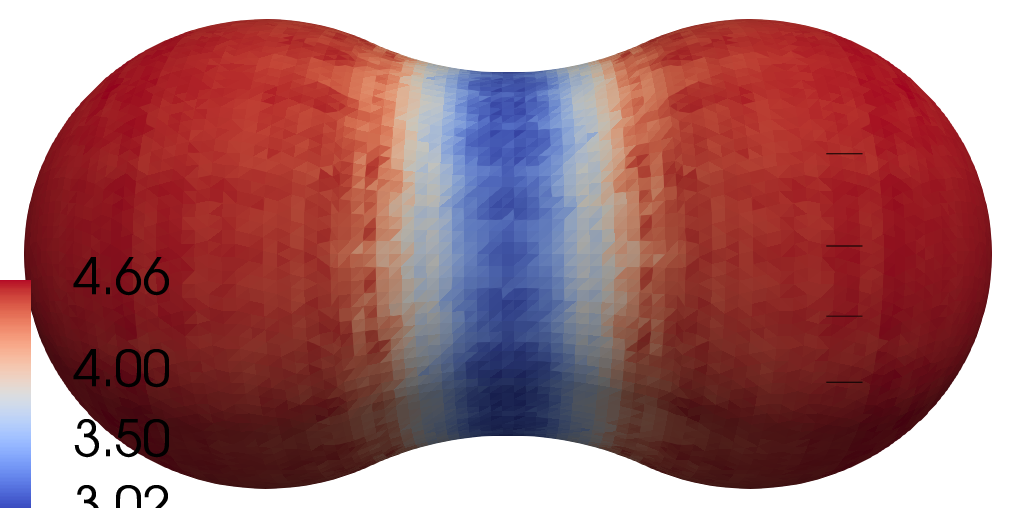}
\caption{Surface potential (left) and magnitude of the surface gradient (right) recovered from the potential field (\ref{eqn:exact_field}) in $\Omega_s$ for 
diatomic molecule using PPR (top row) and enriched recovery (bottom). The nearest atom is used in the enrichment. Mesh size $h=0.2$\AA.}
\label{fig:2atm_exact}
\end{center}
\end{figure}

Our diatomic model consists of two unit spheres centered respectively at $(\pm 1,0.0)$, each carrying a unit positive charge at its center.
We first set an analytical potential field (\ref{eqn:exact_field}) in $\Omega_s$ and recover the potential and potential gradient on surface using PPR and enriched
recovery. If both charges are included for enrichment then recovered results will be accurate with machine error. Instead we choose the charge nearest to a surface 
point for the recovery there. It is seen from Table (\ref{tab:2atm_exact_compare}) that the enrichment with a single Green's function significantly improves the accuracy
of the gradient recovery. At $h=0.4$\AA.~the relative error in surface gradient from enriched recovery is less than $8\%$ while the traditional PPR method 
gives a surface gradient with an relative error more than $80\%$. This significant difference is also highlighted in Fig. \ref{fig:2atm_exact}, where a smoother 
and symmetric surface gradient is delivered by the enriched recovery.

\begin{table}[!ht]
\begin{center}
	\begin{tabular}{ r llllcllll } \hline
                \multirow{2}{*}{$h$} & \multicolumn{4}{c}{PPR} &  & \multicolumn{4}{c}{Enriched Recovery} \\  \cline{2-5} \cline{7-10}
		 & $e(\phi)$  & $e_r(\phi)$ & $e(\nabla \phi)$ & $e_r(\nabla \phi)$ & & $e(\phi)$  & $e_r(\phi)$ & $e(\nabla \phi)$ & $e_r(\nabla \phi)$ \\ \hline
		0.4  & 1.35 & 4.80e-2 & 9.60 & 8.17e-1 &  &1.91e-1 & 6.44e-3 & 1.46 & 7.78e-2 \\
		0.2  & 2.77e-1 & 9.51e-3 & 3.98 & 2.31e-1  & &  1.73e-2 & 5.82e-4 & 2.68e-1 & 1.24e-2 \\ 
		0.1  & 5.57e-2 & 1.90e-3 & 1.39 & 7.08e-2  & & 3.49e-3 & 1.18e-4 & 8.90e-2 & 4.13e-3 \\
		0.05 & 9.12e-3 & 3.11e-4 & 4.23e-1 & 2.06e-2 & & 5.48e-4 & 1.85e-5 & 2.69e-2 & 1.25e-3 \\ \hline	
	\end{tabular}
	\caption{Errors and relative errors in surface potential and gradient recovered using PPR and enriched recovery from the 
potential field (\ref{eqn:exact_field}) in $\Omega_s$ for the diatomic molecule. The nearest atom is used in the enrichment.}
\label{tab:2atm_exact_compare}
\end{center}
\end{table}

\begin{figure}[!ht]
\begin{center}
\includegraphics[height=3cm]{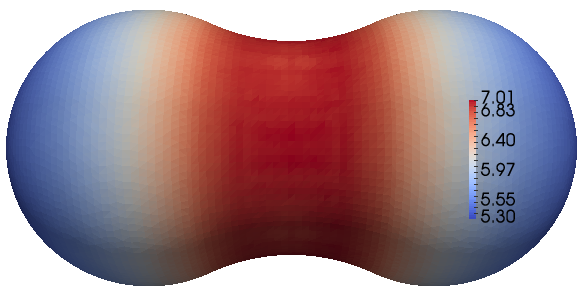} \hspace{1cm}
\includegraphics[height=3cm]{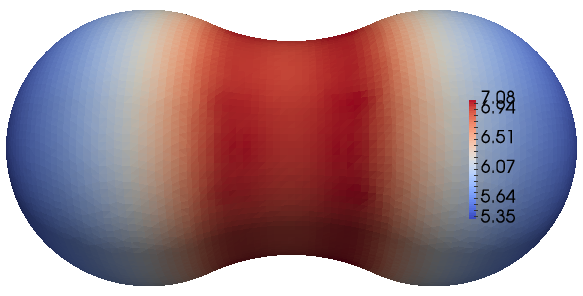} \\
\includegraphics[height=3cm]{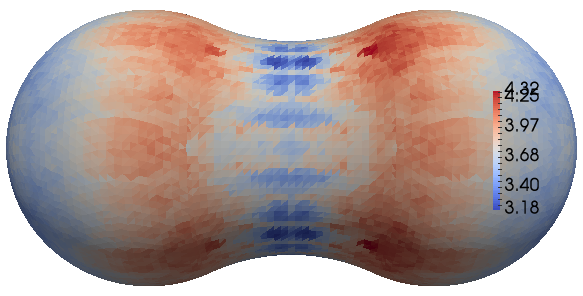} \hspace{1cm}
\includegraphics[height=3cm]{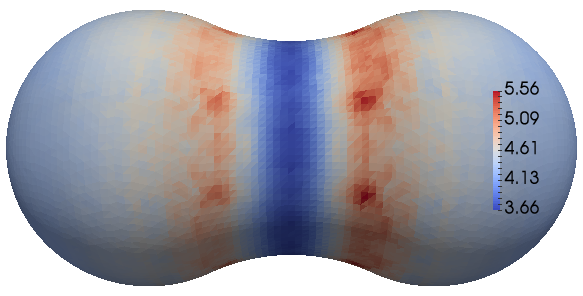} \\
\includegraphics[height=3cm]{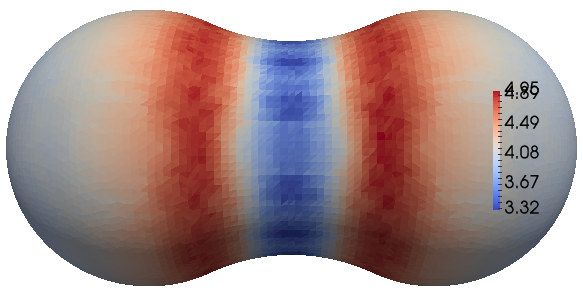} \hspace{1cm}
\includegraphics[height=3cm]{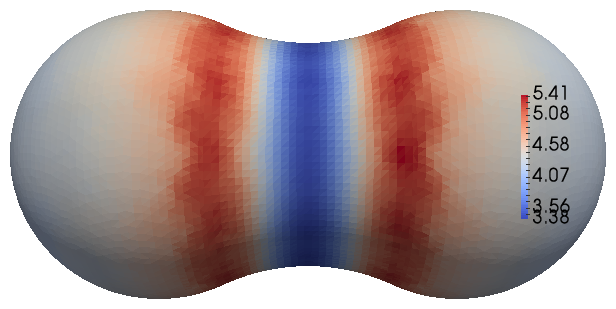} \\
\includegraphics[height=3cm]{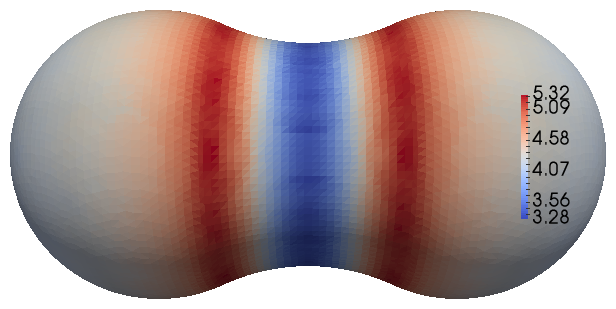} \hspace{1cm}
\includegraphics[height=3cm]{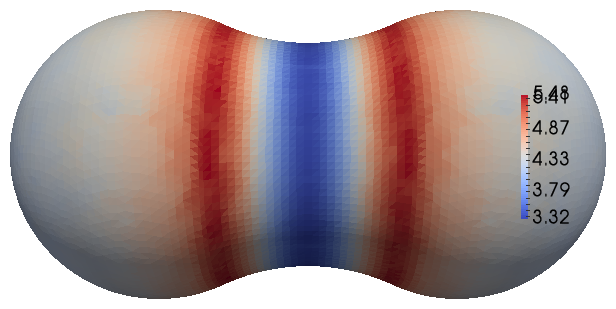} \\
\caption{Surface potential (top row) and surface gradient (rest) recovered using PPR (left) and enriched recovery (right) from the numerical solutions of the 
PBE for solvated diatomic molecule. Both atoms are used in the enrichment. Mesh size $h=0.2,0.2,0.1$ and $0.05$\AA~from top to bottom.}
\label{fig:2atm_mib_grd_compare}
\end{center}
\end{figure}

Th enriched recovery is also tested and compared to PPR on the numerical solutions of the PBE for this diatomic molecule. Since there does not exist analytical 
values of the surface potential and potential gradient, we would use the results of the enriched recovery at a fine mesh as the reference solution for error
measurement. Both atoms are used in the enrichment for optimal recovery. The improvement of the accuracy in the recovered variables due to the enrichment is 
evidenced by the results collected in Table \ref{tab:2atm_mib_compare}. This table and the plots in Fig. \ref{fig:2atm_mib_grd_compare} show that both the
PPR and enriched recovery can deliver accurate surface potential (top row) from the numerical solutions of the PBE but the accuracies of recovered gradients differ
significantly. For all mesh sizes we tested the enriched recovery generates surface gradients with a consistent range. In contrast, the results of PPR vary considerably
as the mesh is refined. At a fine mesh with $h=0.05$\AA~, notable variation is still observed in the surface gradient at the center neck of this diatomic molecule.
The coupled enriched recovery is also tested on the numerical solutions of the PBE and the results are rather similar to the recovery based on the solution in $\Omega_s$
only. 
\begin{table}[!ht]
\begin{center}
	\begin{tabular}{ r llllcllll } \hline
                \multirow{2}{*}{$h$} & \multicolumn{4}{c}{PPR} &  & \multicolumn{4}{c}{Enriched Recovery} \\  \cline{2-5} \cline{7-10}
		 & $e(\phi)$  & $e_r(\phi)$ & $e(\nabla \phi)$ & $e_r(\nabla \phi)$ & & $e(\phi)$  & $e_r(\phi)$ & $e(\nabla \phi)$ & $e_r(\nabla \phi)$ \\ \hline
		0.4  &  5.46e-1 & 1.86e-2 & 9.51 & 7.72e-1 & & 8.23e-1    & 3.43e-2 & 1.08 & 7.22e-2 \\
		0.2  &  1.86e-1 & 6.32e-3 & 4.37 & 2.53e-1 & & 2.76e-1 & 9.54e-3 & 4.72e-1 & 2.05e-2 \\
		0.1  &  5.71e-2 & 1.94e-3 & 1.54 & 7.74e-2 & & 2.97e-2 & 9.89e-4 & 2.09e-1 & 9.18e-3 \\
		0.05 &  9.76e-3 & 3.31e-4 & 4.58e-1 & 2.20e-2 & &   &  &   & \\ \hline
	\end{tabular}
	\caption{Errors and relative errors in surface potential and gradient recovered using PPR and enriched recovery from the numerical solutions of the PBE
for solvated diatomic molecule. Both atoms are used in the enrichment. Errors are measured using the enriched recovery at $h=0.05$\AA~as the reference solutions.}
\label{tab:2atm_mib_compare}
\end{center}
\end{table}

\subsubsection{Amino Acids} \label{sect:residues}
The numerical experiments above show that the accuracy of recovered gradients can be increased by enriching Green's functions at all charged atoms. For real biomolecular
simulations we might not be able to afford this global enrichment because of the increasing computational cost. Global enrichment might not be necessary because the Green's functions
decay in $1/r$, which indicates the potential is prone to be dominated by the charges nearby. It is therefore possible only to choose a small set of charged atoms
close to the surface point of recovery. Here we will test the enriched recovery on six amino acids: TRY, ASN, ASP, ARG, LYS and GLU dimer. For each amino acid, we set an analytical 
potential field using (\ref{eqn:exact_field}) assuming a unit positive charge at each atom.  

As seen in Tables (\ref{tab:TRY_exact},\ref{tab:ASN_exact}), the classical PPR ($m=0$) has to use a very fine mesh $h=0.05$\AA to generate a surface gradient with a relative error 
about $2\%$. Recovery enriched with Green's functions at $5$ nearest atoms is able to deliver results of the similar accuracy at $h=0.4$\AA. This number is also identified on 
the results for other residues, which are neglected as they show the very similar behavior as in these two tables. Enrichment with more Green's functions will further improve
the accuracy of the recovery but for the cause of efficiency we will choose $m=5$ in the numerical experiments below.
\begin{table}[!ht]
\begin{center}
	\begin{tabular}{ r llcllcll } \hline
                \multirow{2}{*}{$m$} & \multicolumn{2}{c}{$h=0.4$} & & \multicolumn{2}{c}{$h=0.2$} & & \multicolumn{2}{c}{$h=0.05$}\\  \cline{2-3} \cline{5-6} \cline{8-9}
    &  $e_r(\phi)$ & $e_r(\nabla \phi)$ & & $e_r(\phi)$ & $e_r(\nabla \phi)$ & & $e_r(\phi)$ & $e_r(\nabla \phi)$ \\ \hline
	0  & 1.16e-1 & 8.00e-1  & &  2.53e-2 & 2.46e-1 & & 7.12e-4 & 2.10e-2 \\ 
	1  & 3.66e-2 & 1.62e-1  & &  7.06e-3 & 5.68e-2 & & 1.86e-4 & 5.49e-3 \\
	3  & 1.05e-2 & 4.35e-2  & &  1.41e-3 & 1.17e-2 & & 3.87e-5 & 1.17e-3 \\
	5  & 5.20e-3 & 2.22e-2  & &  6.87e-4 & 5.44e-3 & & 2.00e-5 & 5.71e-4 \\
        8  & 2.28e-3 & 9.67e-3  & &  1.81e-4 & 1.47e-3 & & 2.02e-5 & 5.19e-4 \\
       11  & 8.80e-4 & 3.62e-3  & &  5.90e-5 & 4.99e-4 & & 5.70e-6 & 1.39e-4 \\
       15  & 3.13e-4 & 1.35e-3  & &  1.49e-5 & 1.23e-4 & & 1.08e-6 & 2.84e-5 \\
       20  & 1.70e-5 & 9.04e-5  & &  5.80e-7 & 5.17e-6 & & 5.25e-7 & 1.32e-5 \\ \hline
\end{tabular}
	\caption{Relative errors in the potential and gradient obtained from recovery enriched with different number of Green's functions $m$ for TYR surface.  
Exact potential field (\ref{eqn:exact_field}) is used. TYR consists of $21$ atoms.}
\label{tab:TRY_exact}
\end{center}
\end{table}

\begin{table}[!ht]
\begin{center}
	\begin{tabular}{ r llcllcll } \hline
                \multirow{2}{*}{$m$} & \multicolumn{2}{c}{$h=0.4$} & & \multicolumn{2}{c}{$h=0.2$} & & \multicolumn{2}{c}{$h=0.05$}\\  \cline{2-3} \cline{5-6} \cline{8-9}
    &  $e_r(\phi)$ & $e_r(\nabla \phi)$ & & $e_r(\phi)$ & $e_r(\nabla \phi)$ & & $e_r(\phi)$ & $e_r(\nabla \phi)$ \\ \hline
	0  & 1.05e-1 & 7.89e-1  & &  2.34e-2 & 2.45e-1 & & 6.56e-4 & 2.09e-2 \\
	1  & 3.94e-2 & 1.79e-1  & &  8.25e-3 & 6.94e-2 & & 2.07e-4 & 6.44e-3 \\
	3  & 1.12e-2 & 4.86e-2  & &  1.88e-3 & 1.50e-2 & & 3.63e-5 & 1.16e-3 \\
	5  & 5.20e-3 & 2.22e-2  & &  6.85e-4 & 5.57e-3 & & 1.41e-5 & 4.28e-4 \\
        7  & 2.56e-3 & 1.11e-2  & &  4.27e-4 & 3.19e-3 & & 2.02e-5 & 5.97e-4 \\
        9  & 1.09e-3 & 4.72e-3  & &  1.34e-4 & 9.83e-4 & & 1.03e-5 & 2.89e-4 \\
       12  & 1.80e-4 & 8.52e-4  & &  1.73e-5 & 1.38e-4 & & 2.26e-6 & 6.23e-5 \\ \hline
\end{tabular}
	\caption{Relative errors in the potential and gradient obtained from recovery enriched with different number of Green's functions $m$ for ASN surface. 
Exact potential field (\ref{eqn:exact_field}) is used. ASN consists of $14$ atoms.}
\label{tab:ASN_exact}
\end{center}
\end{table}

\begin{figure}[!ht]
\begin{center}
\includegraphics[height=3cm]{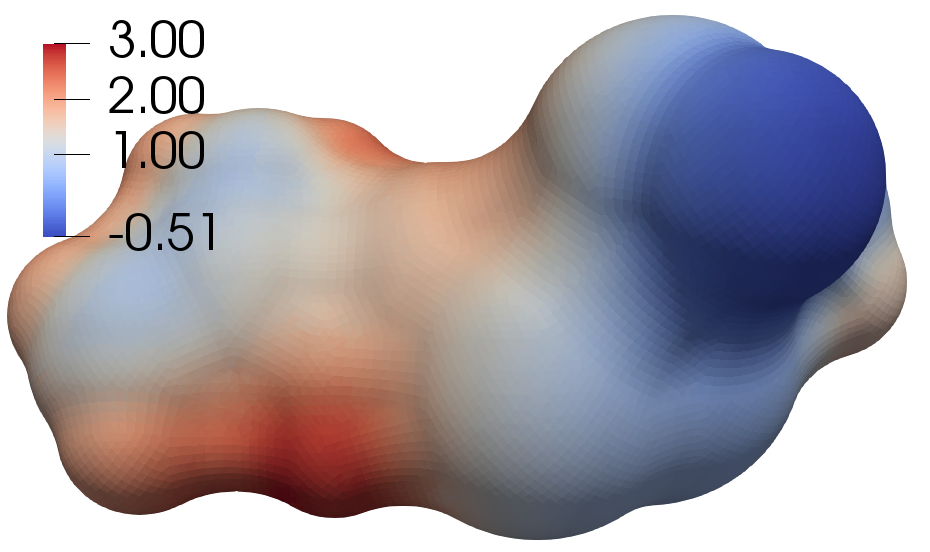}  \hspace{1cm}
\includegraphics[height=3cm]{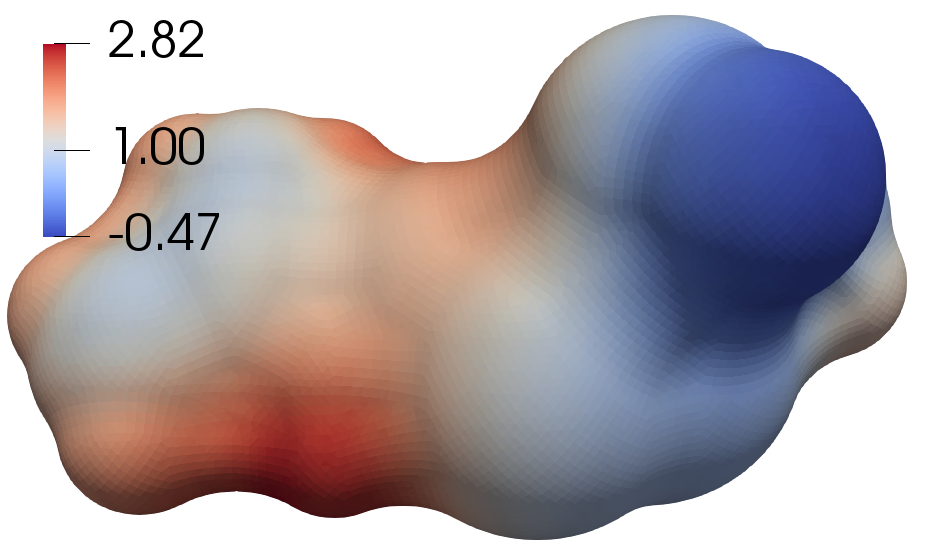} \\
\includegraphics[height=3cm]{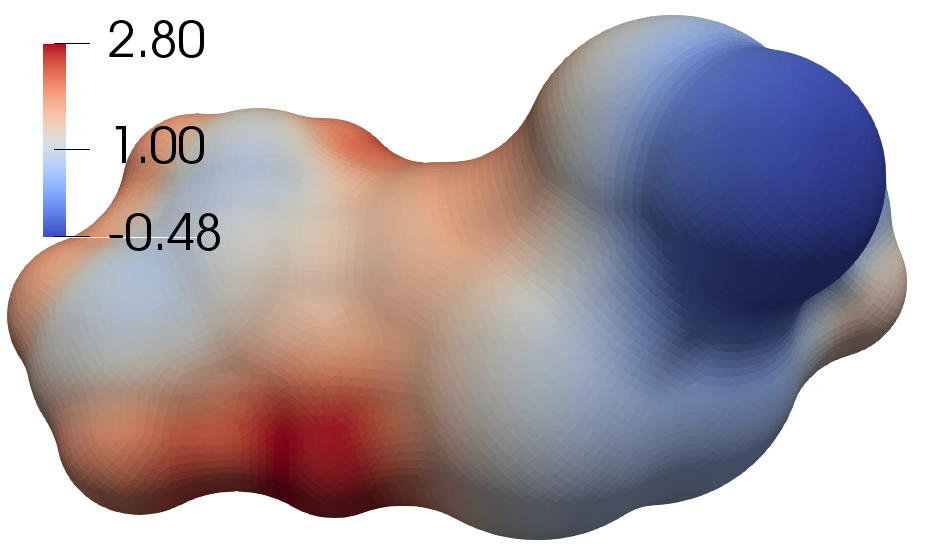}  \hspace{1cm}
\includegraphics[height=3cm]{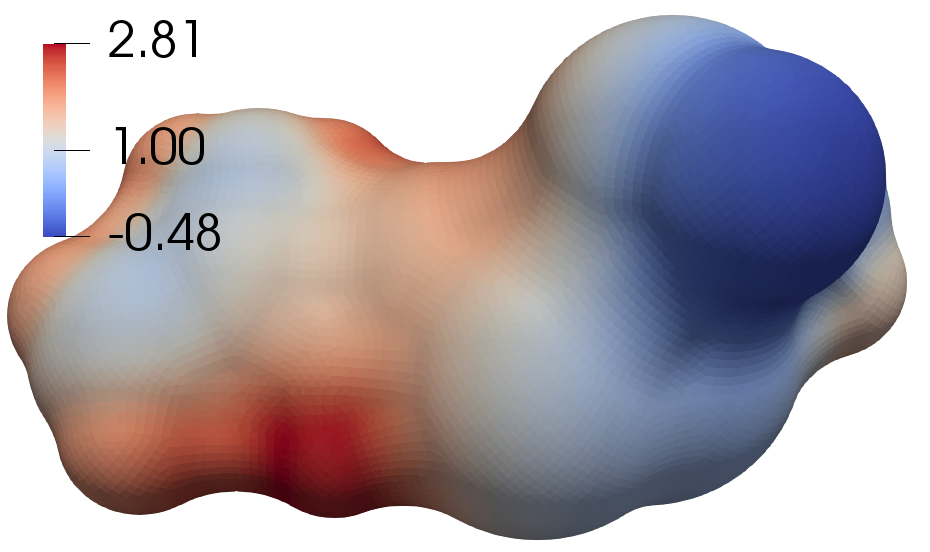} \\
\includegraphics[height=3cm]{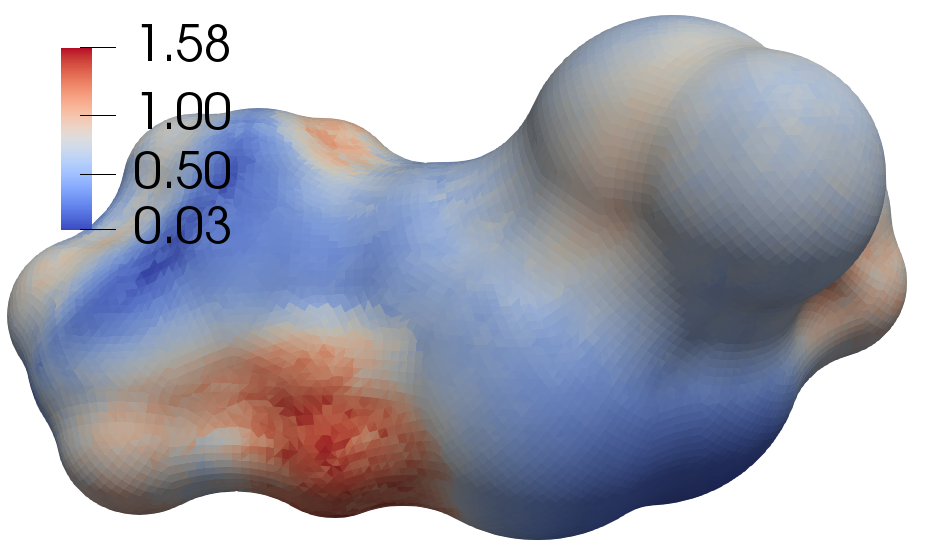}  \hspace{1cm}
\includegraphics[height=3cm]{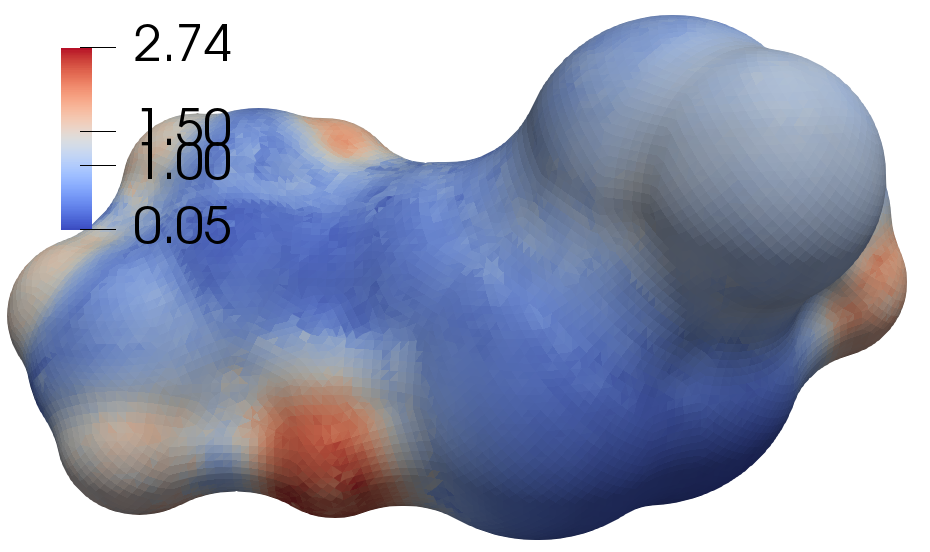} \\
\includegraphics[height=3cm]{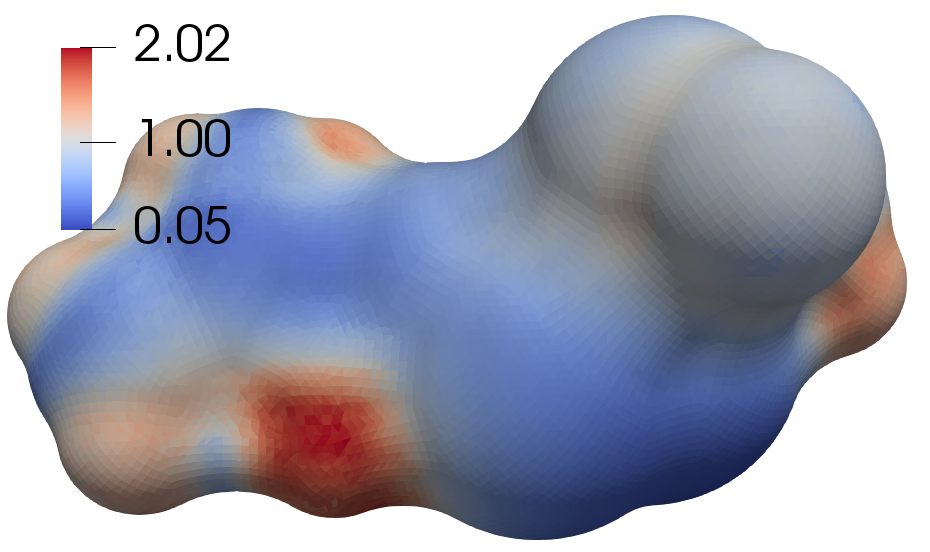}   \hspace{1cm}
\includegraphics[height=3cm]{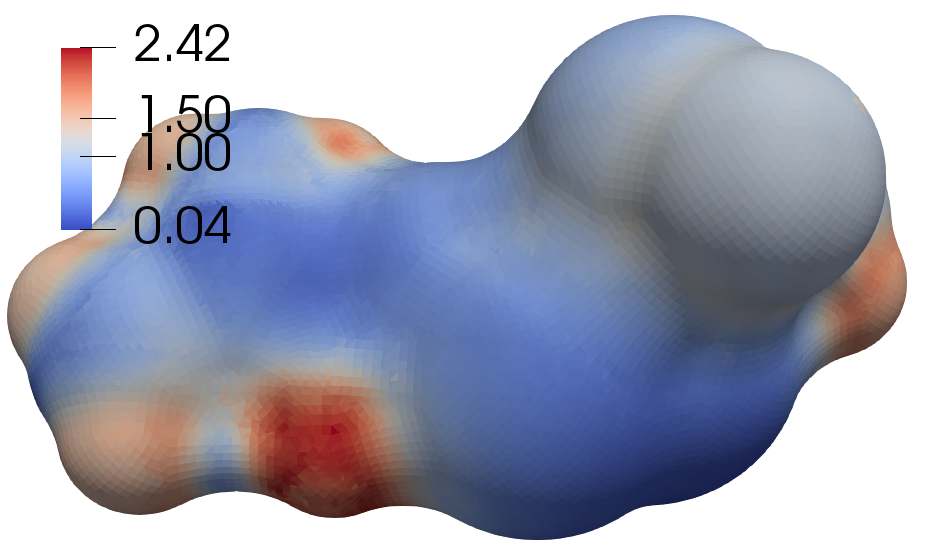}  \\
\includegraphics[height=3cm]{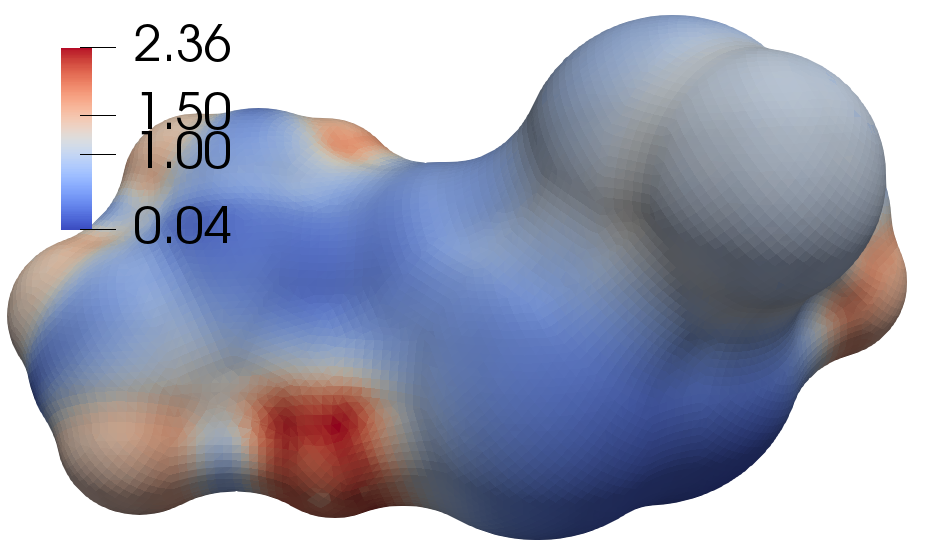}   \hspace{1cm}
\includegraphics[height=3cm]{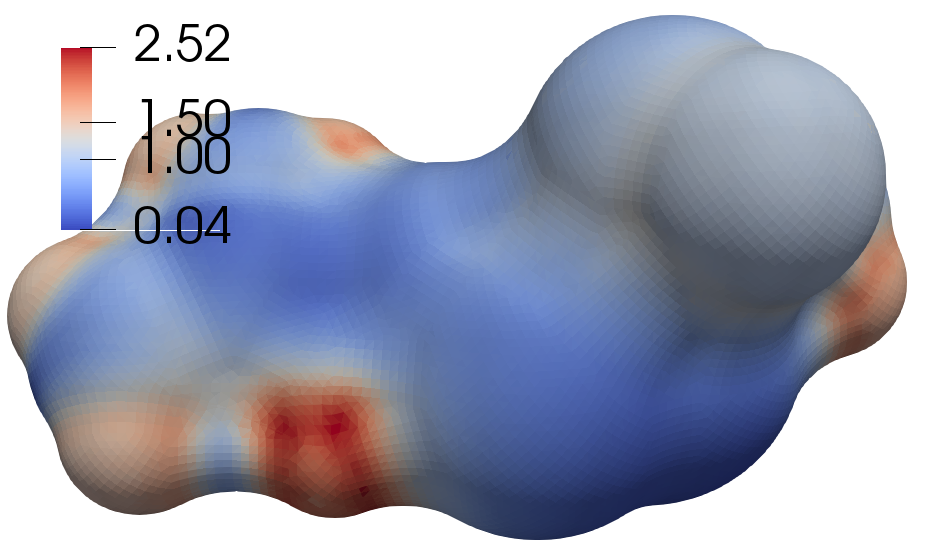}
\caption{Surface potential (top two rows) and magnitude of the surface gradient (others) recovered using PPR (left) and enriched recovery (right) from the numerical solutions of the 
PBE for amino acid ARG. Five closest charged atoms are used in the enrichment. Mesh size $h=0.4,0.1,0.4,0.2$ and $0.1$\AA~from top to bottom.}
\label{fig:ARG_mib_compare}
\end{center}
\end{figure}

\begin{figure}[!ht]
\begin{center}
\includegraphics[height=3cm]{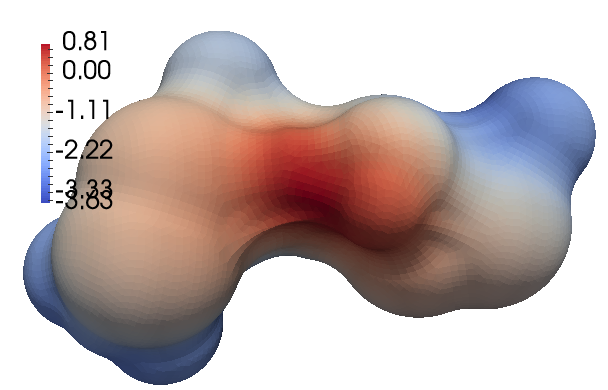}  \hspace{1cm}
\includegraphics[height=3cm]{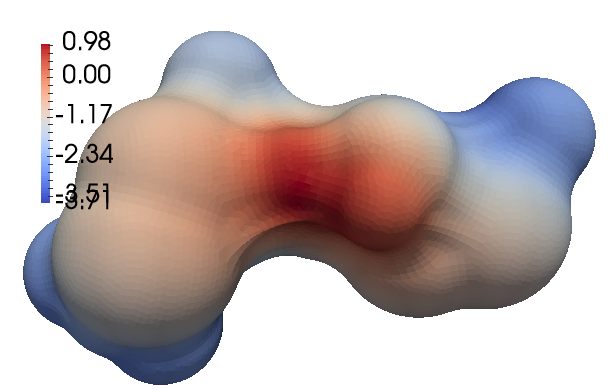} \\
\includegraphics[height=3cm]{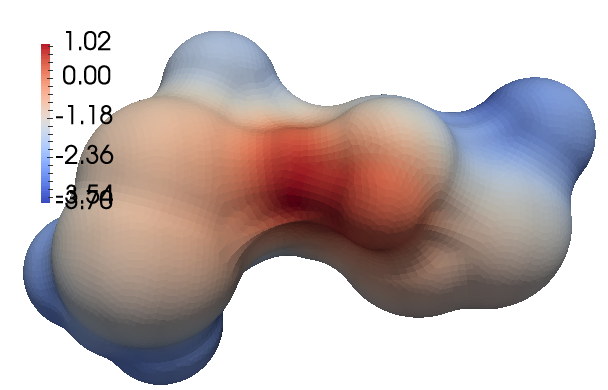}  \hspace{1cm}
\includegraphics[height=3cm]{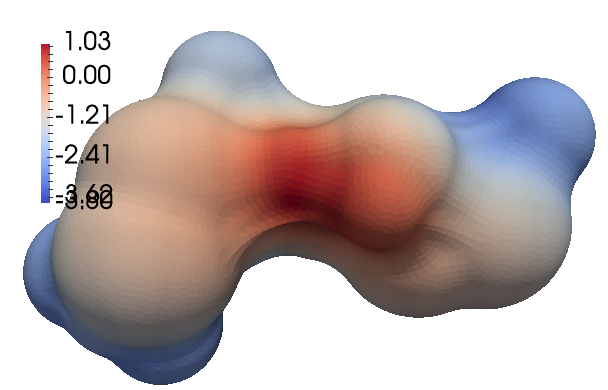} \\
\includegraphics[height=3cm]{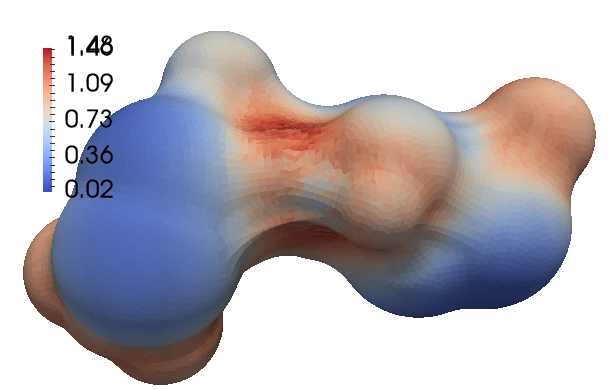}  \hspace{1cm}
\includegraphics[height=3cm]{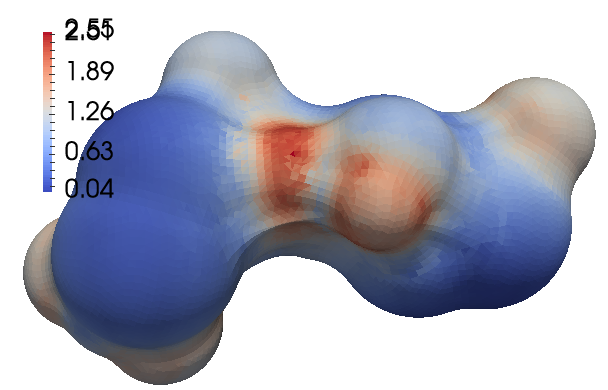} \\
\includegraphics[height=3cm]{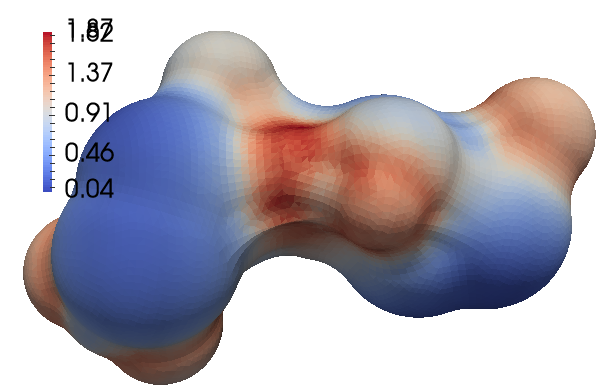}   \hspace{1cm}
\includegraphics[height=3cm]{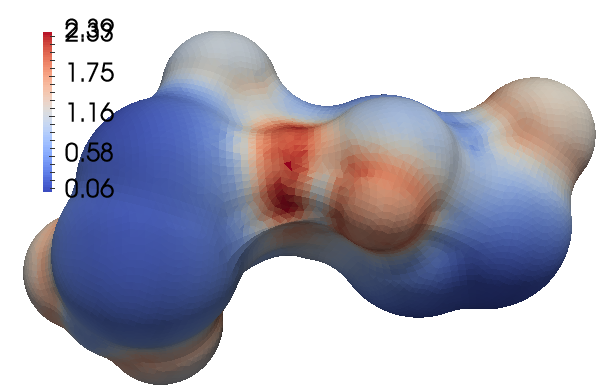}  \\
\includegraphics[height=3cm]{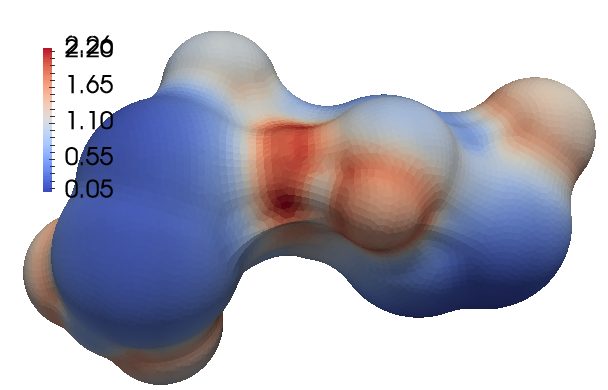}   \hspace{1cm}
\includegraphics[height=3cm]{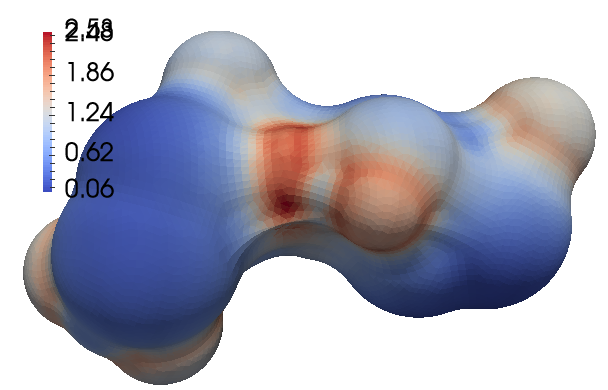}
\caption{Surface potential (top two rows) and magnitude of the surface gradient (others) recovered using PPR (left) and enriched recovery (right) from the numerical solutions of the 
PBE for a GLU dimer. Five closest charged atoms are used in the enrichment. Mesh size $h=0.4,0.1,0.4,0.2$ and $0.1$\AA~from top to bottom.}
\label{fig:GLU2_mib_compare}
\end{center}
\end{figure}

We shall also compare PPR with enriched recovery on numerical solutions of the PBE for these amino acids. The surface potential and gradients recovered with the global 
enrichment at $h=0.05$\AA~ are taken as the reference solutions to examine the accuracy of recovery scheme enriched with different number of Green's function and 
different $h$. The plots for ARG in Fig. \ref{fig:ARG_mib_compare} and for GLU dimer 
in Fig. \ref{fig:GLU2_mib_compare} show that surface potential recovered by the classical PPR has an error more than $10\%$ with respect to enriched recovery 
on a coarse mesh with $h=0.4$\AA; the surface gradient delivered by PPR is indeed errant on the central neck of the GLU dimer. Despite the salient errors in the surface quantities 
recovered by PPR on coarse meshes, these results converge to the same limit as the enriched recovery when the mesh of the numerical solutions for the PBE is sufficiently refined. 
In contrast, the enriched recovery produces very consistent surface potentials and surface gradients on all meshes tested, c.f., the 
ranges of these surface quantities in these plots. 

\subsubsection{Proteins} \label{sect:proteins}
\begin{figure}[!ht]
\begin{center}
\includegraphics[height=3cm]{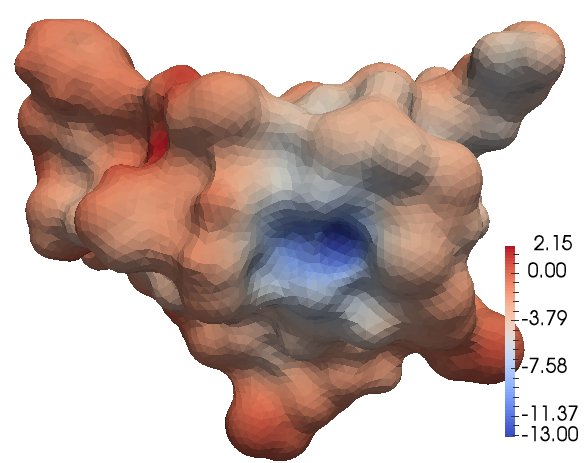}  \hspace{1cm}
\includegraphics[height=3cm]{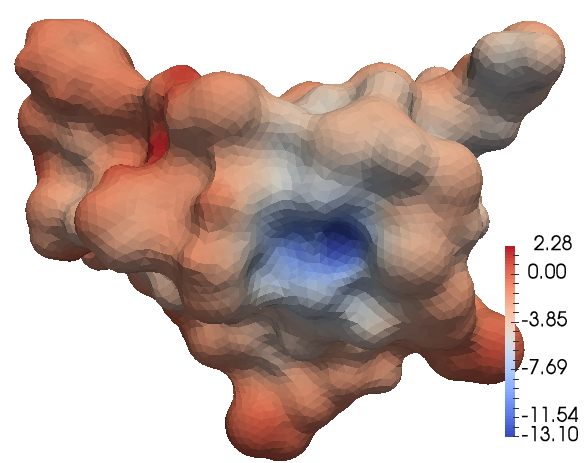} \\
\includegraphics[height=3cm]{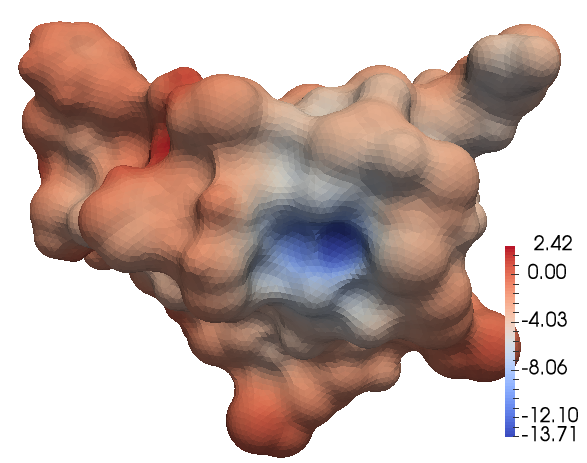} \hspace{1cm}
\includegraphics[height=3cm]{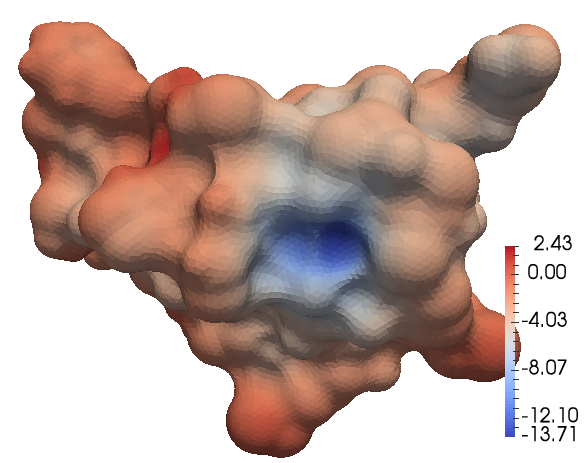} \\
\includegraphics[height=3cm]{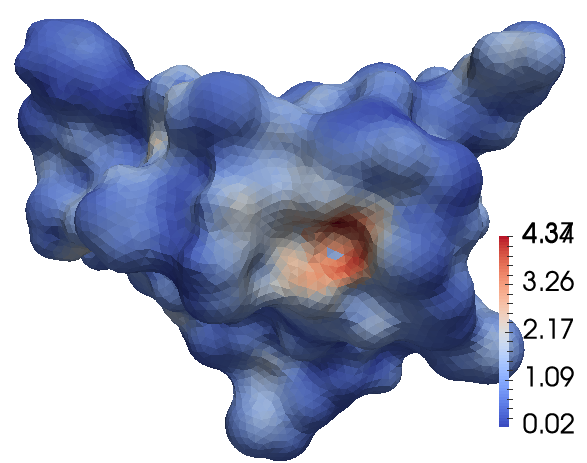} \hspace{1cm}
\includegraphics[height=3cm]{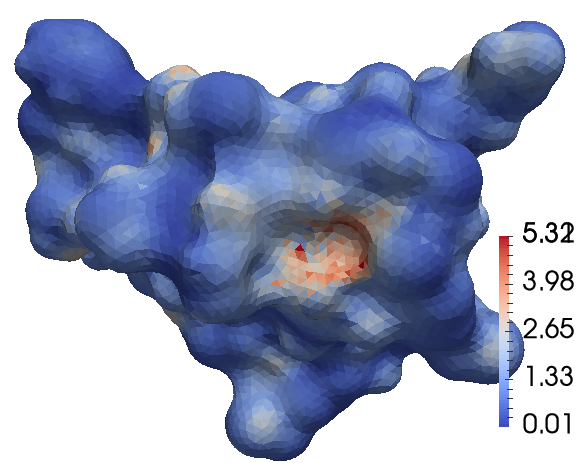} \\
\includegraphics[height=3cm]{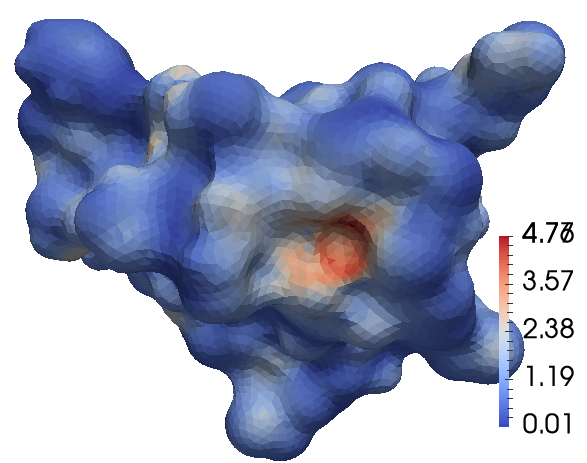} \hspace{1cm}
\includegraphics[height=3cm]{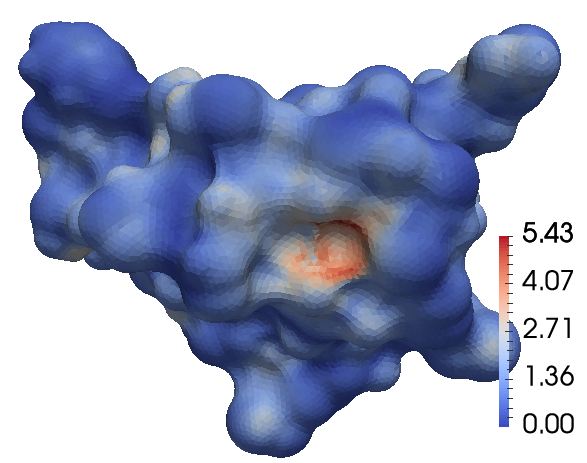} \\
\includegraphics[height=3cm]{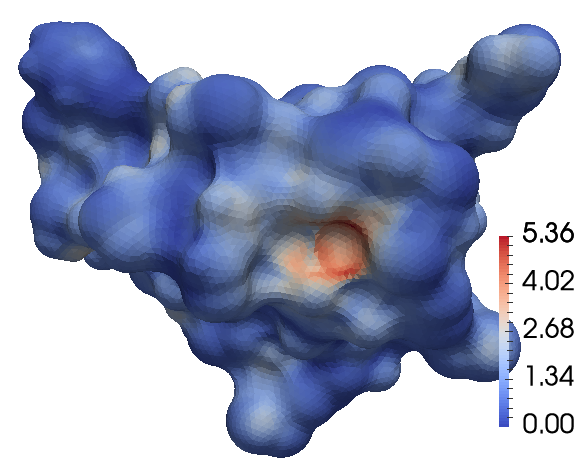} \hspace{1cm}
\includegraphics[height=3cm]{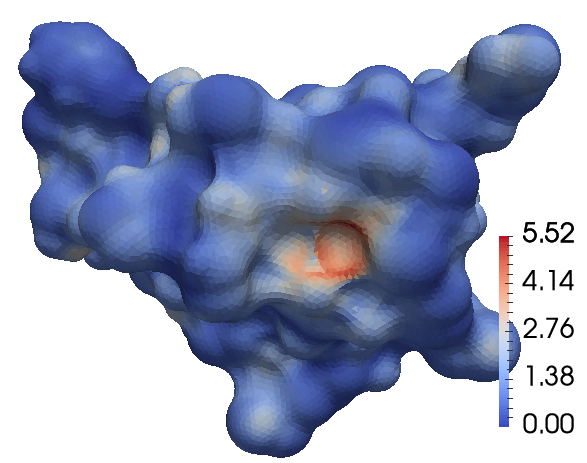}
\caption{Surface potential (top two rows) and magnitude of the surface gradient (others) recovered using PPR (left) and enriched recovery (right) from the numerical solutions of the 
PBE for solvated protein (PDB ID: 1AJJ). At most five closest charged atoms with a distance of $8$\AA~are used in the enrichment. Mesh size $h=0.4,0.1,0.4,0.2$ and $0.1$\AA~from top to bottom.}
\label{fig:1ajj_mib_compare}
\end{center}
\end{figure}

Our last numerical experiments are on a low-density lipoprotein receptor 1AJJ which has 37 residues and 519 atoms. The accuracy 
of the recovered surface potential and gradient from the numerical solutions of the PBE is documented in Tab. \ref{tab:1ajj_mib_compare},
where the results of the enriched recovery on the uniform mesh with $h=0.05$\AA~is used as the reference. This table provides evidence of the 
higher accuracy of the enriched recovery over the classical PPR. For $h=0.4$\AA, a relative error over $40\%$ is found in the surface gradient recovered by PPR. 
This error is reduced to about $4\%$ on a very fine mesh with $h=0.1$\AA, which is still larger than the enriched recovery on a coarse mesh with 
$h=0.4$\AA. The accuracy improvement due to the enrichment is also indicated in Fig. \ref{fig:1ajj_mib_compare}. Consistent with the observations on model 
molecules and amino acids above, enriched recovery produces more stable ranges of surface potential and gradient than PPR.  

\begin{table}[!ht]
\begin{center}
	\begin{tabular}{ r llllcllll } \hline
                \multirow{2}{*}{$h$} & \multicolumn{4}{c}{PPR} &  & \multicolumn{4}{c}{Enriched Recovery} \\  \cline{2-5} \cline{7-10}
		 & $e(\phi)$  & $e_r(\phi)$ & $e(\nabla \phi)$ & $e_r(\nabla \phi)$ & & $e(\phi)$  & $e_r(\phi)$ & $e(\nabla \phi)$ & $e_r(\nabla \phi)$ \\ \hline
		0.4  &  1.61 & 1.87e-2 & 11.0 & 4.65e-1 & & 1.52e-1 & 1.76e-3 & 9.49e-1 & 3.03e-2 \\
		0.2  &  3.59e-1 & 4.15e-3 & 4.50 & 1.60e-1 & & 1.73e-2 & 2.00e-4 & 2.18e-1 & 6.96e-3 \\
		0.1  &  6.31e-2 & 7.30e-4 & 1.48 & 4.90e-2 & & 2.70e-3 & 3.13e-5 & 6.28e-2 & 2.01e-3 \\ \hline
	\end{tabular}
	\caption{Errors and relative errors in surface potential and gradient recovered using PPR and enriched recovery from the numerical solutions of the PBE
for solvated protein (PDB ID: 1AJJ). At most five closest charged atoms within a distance of $8$\AA~are used in the enrichment.}
\label{tab:1ajj_mib_compare}
\end{center}
\end{table}

\section{Concluding Remarks} \label{sect:summary}
We present in this paper a novel numerical method for computing the potential and its gradient on the molecular surfaces from the numerical solutions
of the Poisson-Boltzmann equation, which is widely used in modeling the electrostatic interactions for biomolecules in solvent. Our method reconstructs
a solution in the least square sense locally in the solvent region. In contrast to the classical polynomial preserving recovery (PPR) method that is based
only on the polynomial basis functions, our reconstruction enriches the basis with Green's functions modeling the electrostatic potential field induced 
by selected charges. This enrichment is motivated by the Generalized Born method which decomposes the electrostatic potential inside the solute molecules
into a Coulomb potential and a reaction field, and also by the approximate boundary condition for the Poisson-Boltzmann equation. Extensive numerical methods 
on Born atom, model diatomic molecule, amino acids, and proteins demonstrate that our enriched recovery is more accurate than PPR on the same mesh, more 
stable along the mesh refinement, and indeed provides a stable reference to which the PPR solutions will converge.

Our enriched recovery is easy to implement, and can be readily integrated with other interface finite difference or finite element methods for the 
Poisson-Boltzmann equation. As the reconstruction is independently computed at each surface point of interest, parallelization of this enriched recovery is
straightforward. We are currently working to supply the recovered surface gradient to compute the dielectric boundary force and use the force to 
drive the large scale continuum molecular deformation as induced due to varying solvation states such as binding of molecules or ligands, changes in 
ion concentration, or protonation.

\section*{Acknowledgements}
This work has been partially supported by National Institutes of Health through the grant R01GM117593 as 
part of the joint DMS/NIGMS initiative to support research at the interface of the biological and mathematical sciences.



\end{document}